# GPI 2.0: Characterizing Self-Luminous Exoplanets Through Low-Resolution Infrared Spectroscopy

Arlene J. Alemán[a,b], Bruce Macintosh[a,b], Mary Anne Limbach[c], Mark S. Marley[d], Jeffrey K. Chilcote[e], Quinn Konopacky[f], Dmitry Savransky[g]

[a]Department of Physics, Stanford University, 450 Serra Mall, Stanford, CA, USA 94305
[b]Kavli Institute for Particle Astrophysics and Cosmology, 452 Lomita Mall, Stanford, CA, USA 94305
[c]Department of Physics and Astronomy, Texas A&M University, 4242 TAMU, College Station, TX, USA 77843-4242
[d]Department of Planetary Sciences and Lunar and Planetary Laboratory, University of Arizona, 1629 E University Blvd, Tucson, AZ, USA 85721
[e]Department of Physics, University of Notre Dame, 225 Nieuwland Science Hall, Notre Dame, IN, USA 46556
[f]Center for Astrophysics and Space Science, University of California San Diego, 9500 Gilman Dr, La Jolla, CA, USA 92093
[g]Sibley School of Mechanical and Aerospace Engineering, Cornell University, 130 Upson Hall, Ithaca, NY, USA14853

## ABSTRACT

Direct imaging characterization of extrasolar planets is often done at low spectral resolution. We model the spectrograph for the Gemini Planet Imager upgrade (GPI 2.0) and assess the instrument's potential for allowing observers to constrain exoplanet properties through analysis of near-infrared spectra. We simulated noisy observations followed by calculations of posterior distributions from maximum likelihood comparison with the Sonora 2018 model grid. Preliminary results suggest that GPI 2.0 should allow observers to constrain temperature with sufficient accuracy, but gravity remains largely uncertain. We also explore the effects of incorporating convolution with the instrument line spread function into our simulation and compare the results with our preliminary findings.

## 1. INTRODUCTION

The field of exoplanet science is unique in the sense that observations of the objects of interest have been historically dominated by indirect techniques. Constituting almost 97% of exoplanet discoveries, the transit and radial velocity methods have monopolized exoplanet detection for decades. Advancements in direct-imaging instrumentation and techniques, however, have made direct imaging a powerful contender in recent years. While transit and radial velocity techniques generally allow astronomers to measure a planet's orbit, radius, and/or mass, direct spectroscopy allows us to measure the temperature, gravity, and atmospheric composition of an exoplanet. Studying such features can provide insight into a planet's formation history, including its origin, migration pathways, and the fraction of gas and solids it accreted throughout its lifetime.

### 1.1 The Gemini Planet Imager

Many instruments optimized for direct detection of extrasolar planets use integral field spectrographs (IFSs). The Gemini Planet Imager (GPI) is a coronagraphic adaptive optics integral field spectrograph designed for spectroscopy of young, Jupiter-mass extrasolar planets. If these are optimized for a wide field of view (e.g., to discover previously unknown planets or to better measure the halo of scattered light), the instruments will necessarily be of low spectral resolution to fit within the finite number of detector pixels. GPI's IFS has 200 x 200 spatial samples, but each individual spectrum is only 16 pixels long, typically sampling one of the standard infrared bands. The Gemini Planet Imager Exoplanet Survey (GPIES, Nielsen et al., 2019[1]), observed its 531st and final new star in January 2019. Following this feat, the instrument has been decommissioned in preparation for its relocation from Gemini South to Gemini North and will undergo a series of upgrades aimed at improving its detection capabilities. The upgraded instrument, entitled "GPI 2.0," will include a new wavefront sensor for observations of faint stars (1~14 magnitude), a faster real-time computer for higher performance on bright stars ($10^{-7}$ contrast at 0.5"), greater automation for guest investigator programs and large surveys, and new coronagraph masks.

When designing a new instrument or planning an upgrade of an existing one, it is critical to match instrumental properties (such as spectral resolution) to scientific requirements (such as characterizing exoplanets.) This paper primarily focuses on the upgrades to the integral field spectrograph which includes new spectroscopic prisms that will allow for new observing capabilities (Limbach et al., 2020[2]). GPI's spectrograph operates in the near-infrared wavelength range of 1-2.5 microns and the original GPI design carried 5 filters: Y, J, H, K1, and K2. The prism was designed for a spectral resolution of R ~ 50 in H-band, however, due to the dispersion of the glasses used, resolution varies significantly with wavelength, resulting in the split of K-band into the sub-bands K1 and K2.

Table 1. Existing GPI: SFTM-16/BaF2

| Band | Wavelength Range | Center Resolution |
| --- | --- | --- |
| Y | 0.95-1.14 | 39.17 |
| J | 1.12-1.35 | 35.88 |
| H | 1.5-1.8 | 47.3 |
| K1 | 1.9-2.19 | 72.39 |
| K2 | 2.13-2.4 | 80.02 |

In GPI 2.0, a new prism using P-SF68 glass will have a more uniform resolution and will reunite the two halves of K-band.

Table 2. GPI 2.0: P-SF68/BaF2 Prism

| Band | Wavelength Range | Center Resolution |
| --- | --- | --- |
| Y | 0.95-1.07 | 61.9 |
| J | 1.17-1.33 | 47.9 |
| H | 1.49-1.78 | 42.7 |
| K | 2-2.4 | 55.1 |

Inspired by the Subaru CHARIS spectrograph (Peters-Limbach et al. 2013[3]), GPI 2.0 will also incorporate a broadband mode, capable of simultaneously imaging across Y, J, H, and K. The lower spectral resolutions will enable a higher signal-to-noise ratio on dim stars which may facilitate the tasks of detecting variability and distinguishing planets from background stars and may enhance spectral differential imaging (SDI).

Table 3. GPI 2.0 Low Res.: P-SF68/BaF2 Prism

| Band | Wavelength Range | Resolution |
|---|---|---|
| Y | 0.97-1.07 | 13.6 |
| J | 1.17-1.33 | 10.5 |
| H | 1.49-1.78 | 9.6 |
| K | 2-2.4 | 12.4 |

To optimize these new modes, we simulated GPI spectra of exoplanets and stars spanning a wide range of temperatures. First, we use a simple instrument model to explore the feasibility of constraining temperature and gravity estimates using GPI infrared spectroscopy. This has allowed us to verify that the new prisms will be capable of sufficiently constraining temperature estimates and distinguishing planets from background stars. We also explore optimal strategies for follow-up characterization of planets (i.e., how observations should be allocated between different wavelengths). Secondly, we investigate a more sophisticated model incorporating the effects of the instrument line spread function into our spectrograph simulation and we gauge the significance of this effect on our preliminary findings.

## 2. BASE SIMULATION

### 2.1 Methodology

To explore our research questions, we developed a Python simulation of GPI's spectrograph using the original design parameters of the spectrograph as well as the proposed parameters for its upgrade. In these simulations, we altered various spectrographic properties and investigated the effects that the alterations had on the precision of estimations for certain exoplanet characteristics such as temperature and gravity. Among the instrument and observation parameters we experimented with were spectral resolution, wavelength range, signal-to-noise ratio, and exposure time. A key question to evaluate is the instrument's capacity for distinguishing a legitimate exoplanet signal from that of a background stellar object and what signal-to-noise constraints were required for the instrument to make such a distinction.

Before we could generate observing scenarios and recover likelihood distributions, we needed to establish an appropriate database of spectra to serve as planet models for our simulation. For cooler planets whose temperatures fall between 200K-1600K, we used the Sonora 2018 models of cloudless planets (Marley et al. 2018[4]). For warmer planets and late-type stars, however, the absence of clouds in the Sonora models is somewhat unrealistic. In response, we opted to incorporate real L-type brown dwarf spectra from the Infrared Telescope Facility's SpeX Library to approximate planets (and brown dwarfs) with temperatures between 1600K-2300K.

To properly model the instrument's performance, we first needed to adjust the spectral resolution of the Sonora and SpeX models to realistically represent spectra as they would be observed through GPI or GPI 2.0. To accomplish this, we began by partitioning the total wavelength range into the Y, J, H, and K-bands and the inter-filter (water-band) regions. These partitions were then subdivided into data bins of length Δλ, a value determined based on the spectral resolution of the region. H-band, for example, with its central wavelength at approximately 1.64μm, has a central resolution of R ~ 46.4 which results in a Δλ of roughly 0.035μm. Within each bin, the data points were averaged into a single value that was then assigned to the midpoint of the bin. Henceforth in this publication, "original data files" refers to the reduced-resolution files, unless otherwise specified. Figure 1 below shows an example of what one of the original model spectra would be reduced to at the GPI resolution.

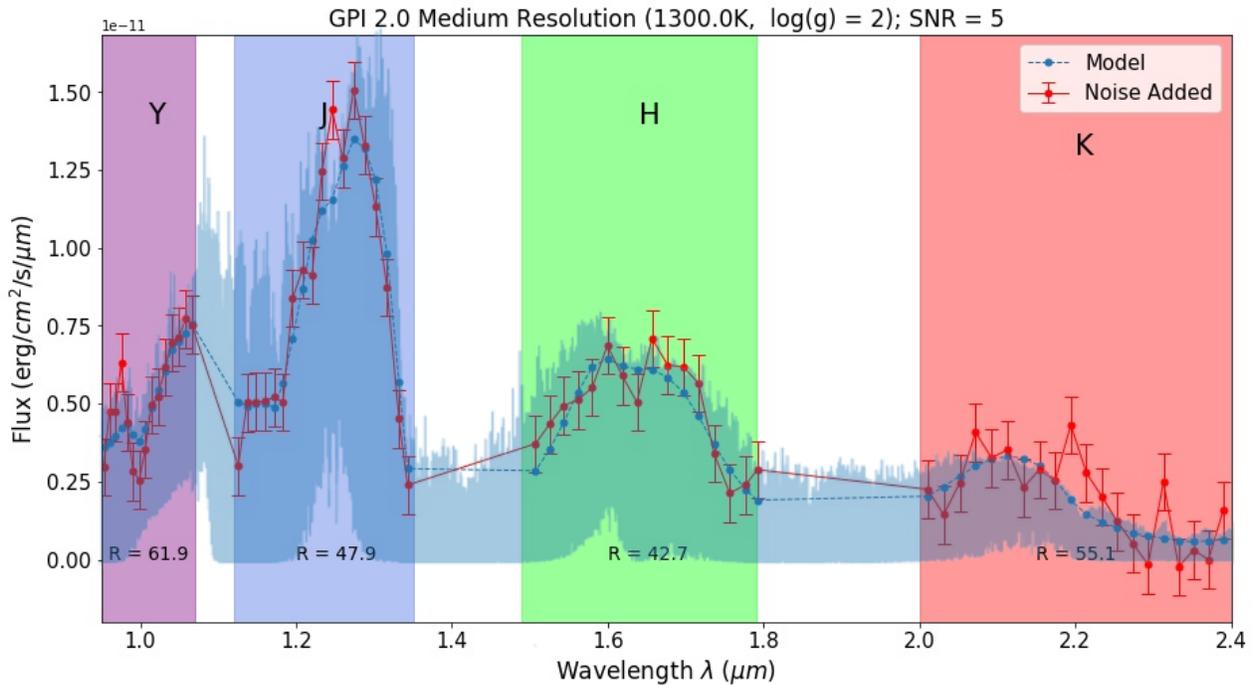

Figure 1. Extrasolar planet spectrum as observed with the upgraded GPI 2.0 prisms. The faint blue, feature-rich spectrum in the background shows the original model resolution. The blue plot overlaid onto it displays the reduced resolution model and the red line displays the reduced resolution model with Gaussian noise at SNR = 5 added.

Having set up a new database of the lower resolution model spectra, the simulation was able to retrieve a given model by searching the database for a match to the specific exoplanet characteristics input by the user. Upon locating the desired model, the subsequent step was to simulate noisy observations by introducing Gaussian noise to the retrieved model. Because of the disproportionate flux across the 4 bands, we initially ran the risk of large signals in certain bands biasing the noise in bands with smaller signals. Consequently, when generating the Gaussian from which we were to pull random numbers, rather than using the mean flux of the full spectrum, we instead used the mean of the flux in H-band which made for a uniform noise profile across the spectrum; i.e., the model achieves the specified SNR averaged over H-band, with the noise having the same absolute (not relative) flux at other wavelengths. With the noisy data from the simulated observation, we proceeded by calculating posterior distributions from maximum likelihood comparison with the grid of models.

## 2.2 Results

We explored variations of a hypothetical observing scenario in which we would have 4 hours to allocate to any combination of filters. We initiated the analysis by allocating all 4 hours of observing time to H-band and followed up by distributing the time between H and the other bands. Because H-band has the best sensitivity to a broad range of planets, GPI uses H-band for initial surveys, so any object discovered with the instrument will always have an H-band image.

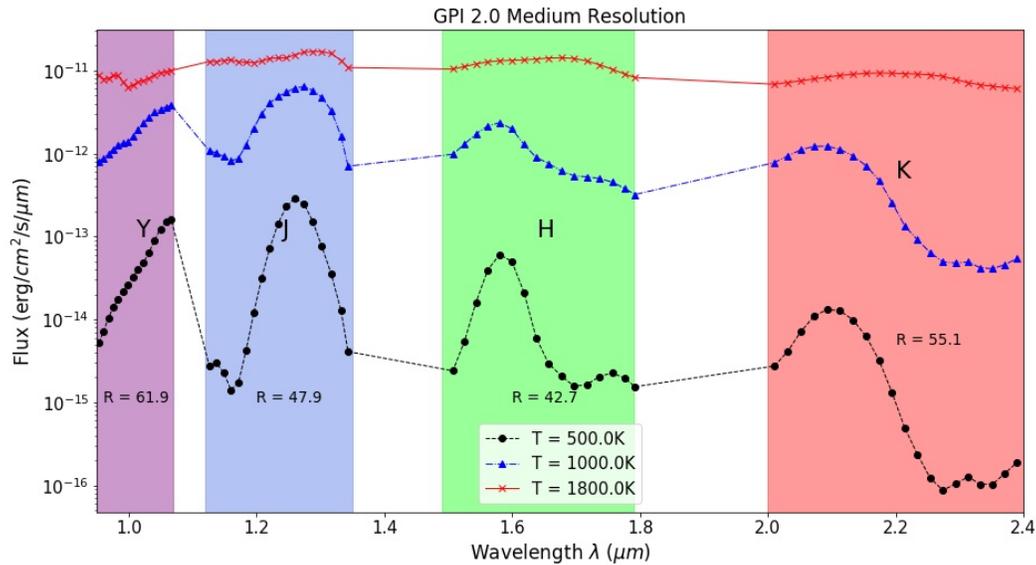

Figure 2. Plot comparing the spectra of planets with different temperatures at GPI 2.0 resolution.

Starting with a simple one-dimensional analysis, we first tested the instrument's proficiency at constraining just one planet parameter. If attempting to constrain temperature, for example, we provided the simulation with the correct gravity value and vice versa. By analyzing the likelihood curves from our one-dimensional cases, it became evident that despite having the capacity to estimate temperature fairly precisely, constraining gravity through GPI spectroscopy would likely prove challenging. Attaining gravity constraints with accuracy comparable to those we were able to acquire for temperature would require a 20-fold SNR improvement.

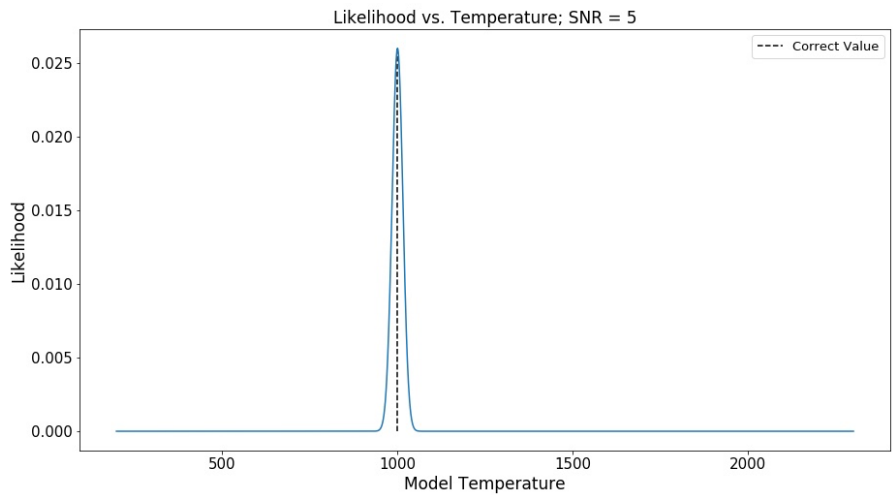

Figure 3. Likelihood plot where the original model temperature was 1000K. The simulated observation used all 4 available filters with 1 hour of exposure time in each and an SNR of 5. The correct gravity value was provided so the simulation was only tasked with constraining temperature.

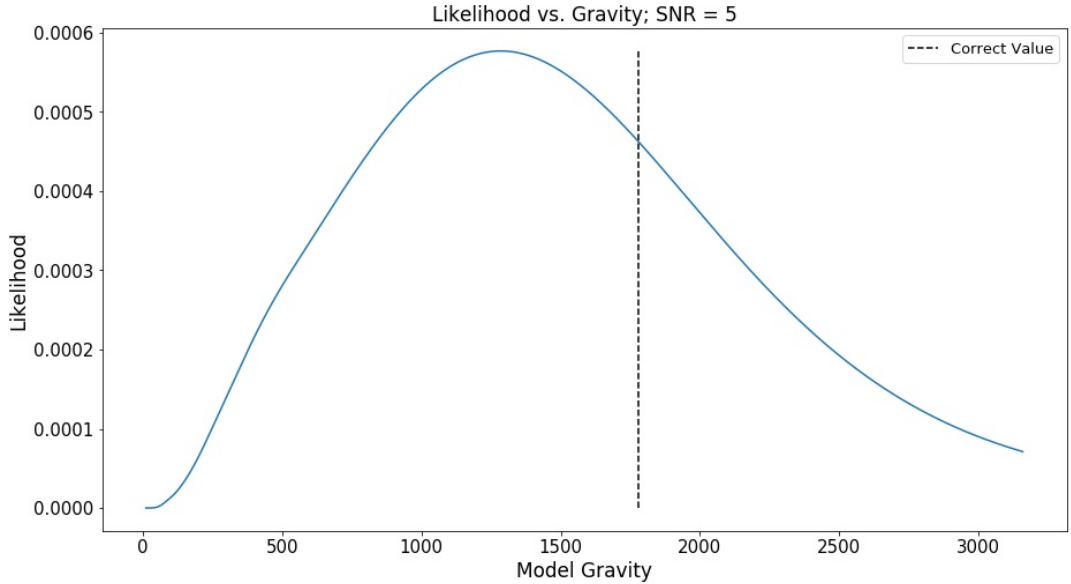

Figure 4. Likelihood plot where the original gravity was 1780 cm/s$^2$ [log(g) = 3.25]. Simulated observation used the same parameters as figure 3's observation. Gravity estimate has significantly more uncertainty than temperature estimate.

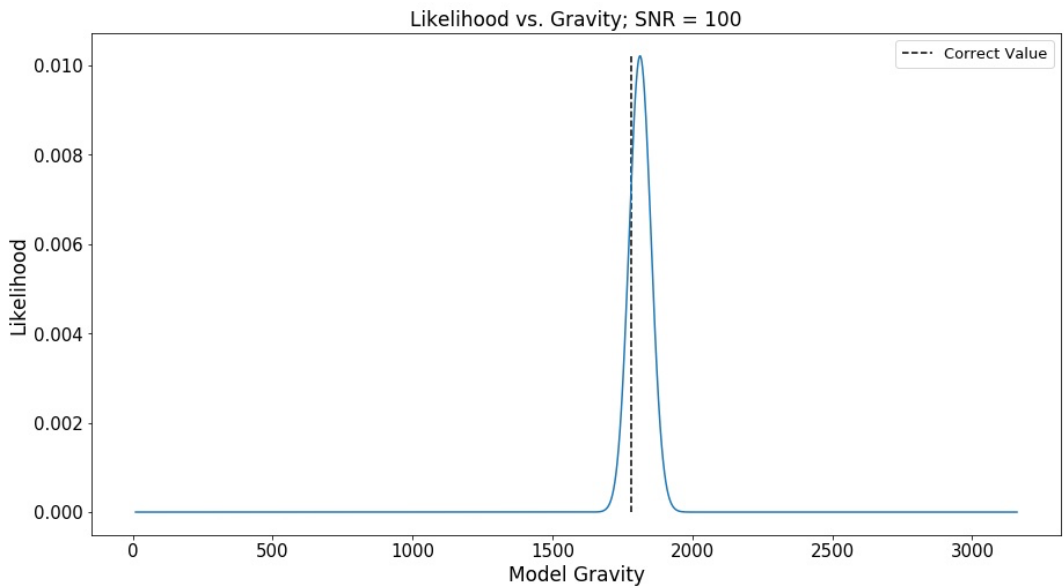

Figure 5. Likelihood plot for the same scenario as figure 4 but with SNR = 100 instead of SNR = 5.

For our two-dimensional analysis, we attempted to constrain temperature and gravity simultaneously without providing the simulation with any prior information about the models' parameters. We recovered promising results for constraining temperature as the simulation continued to demonstrate a high capacity for estimating temperature within a 5-10% margin of error even with relatively low SNR. Additionally, we observed no major differences in the results for models of cooler temperature planets versus hotter temperature planets which suggests little correlation between our model planets' temperatures and the uncertainty in the temperature estimates. Contrastingly, gravity remained largely uncertain and proved even more challenging to estimate without prior knowledge of a model's other characteristics. Figure 6 shows cases where simultaneous gravity and temperature fits are carried out, showing consistently that the temperature is well constrained and gravity (at this low SNR and resolution) remains uncertain.

# 3. LINE SPREAD FUNCTION EFFECTS

## 3.1 Methodology

The methodology for our base simulation – simply binning the model spectrum down to one pixel per instrumental resolution element – is one that is widely implemented when simulating spectrographic instruments. However, this does not exactly reproduce the behavior of a real spectrograph, where optical effects and design choices (such as slit width or lenslet size) blur the spectrum and typically two pixels are required per resolution element. Some cases are more complicated and require more sophisticated modeling. The Roman Space Telescope Coronagraph Instrument (Roman-CGI) is one such example. Due to the non-linear dispersion profile of Roman-CGI's spectroscopic prisms and its broad line spread function (LSF), generating an accurate simulation involves wavelength-to-pixel mapping followed by convolution of the spectra with the instrument LSF. Incorporating these steps enables one to ascertain how much flux lands onto a given detector pixel and to more accurately represent the blurring of spectral features. Carrying out this wavelength-to-pixel mapping and LSF convolution when simulating Roman spectroscopy inspired us to explore how incorporating that methodology into our GPI simulations would affect our earlier results.

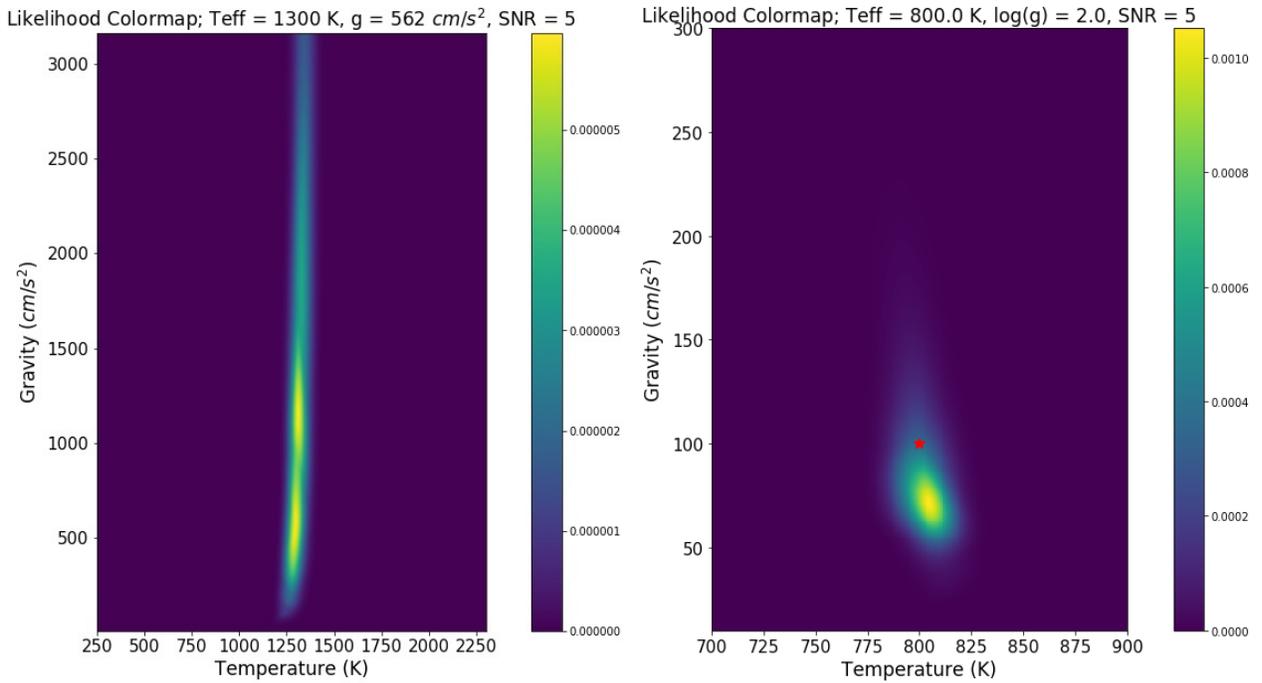

Figure 6. Likelihood colormap displaying results from two different 2-dimensional likelihood tests. At SNR = 5, the simulation was capable of constraining temperature to within 10% of the correct value but suggested that gravity could be almost anything. We were only able to slightly constrain gravity values less than log(g) = 2 as shown in the zoomed in figure on the right.

To develop our new simulation, we started with ZEMAX models mapping a small number of wavelength values to their positions on the spectrograph detector. We had five tables in total, one for each filter and another for the broadband mode. With the information in the tables, we first interpolated in increments of 1μm to get a finer wavelength-to-position mapping.

As the HAWAII-2RG™ (H2RG) detector pixels are 18μm across, we opted to first bin the data into 3μm "subpixels", resulting in 6 subpixels per each 1 pixel. Then similarly to our base simulation, we proceeded by determining which flux values landed into which subpixels, determined the average flux in each subpixel, and assigned the average to the midpoint of the subpixel.

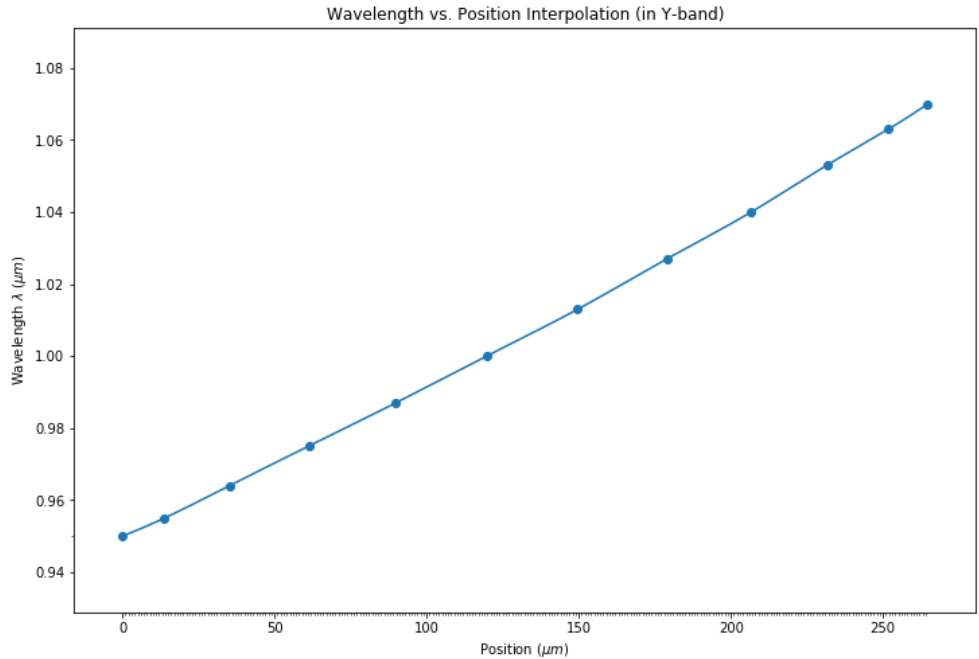

Figure 7. Plot of Y-band wavelength values vs. their positions on the detector interpolated by 1μm increments.

At the subpixel level, we convolved the spectrum with the instrument line spread function which has a full width at half maximum (FWHM) of ~2 pixels for GPI. Since we were convolving at the subpixel level 2 pixels would contain 12 subpixels, so we estimated the LSF with a Gaussian with FWHM ~ 12. Following the convolution step, we averaged the 6 subpixel flux values within each pixel to get the spectra in flux per pixel. With our new database of models, we proceeded by adding Gaussian noise and carrying out maximum likelihood calculations according to our base simulation's methodology.

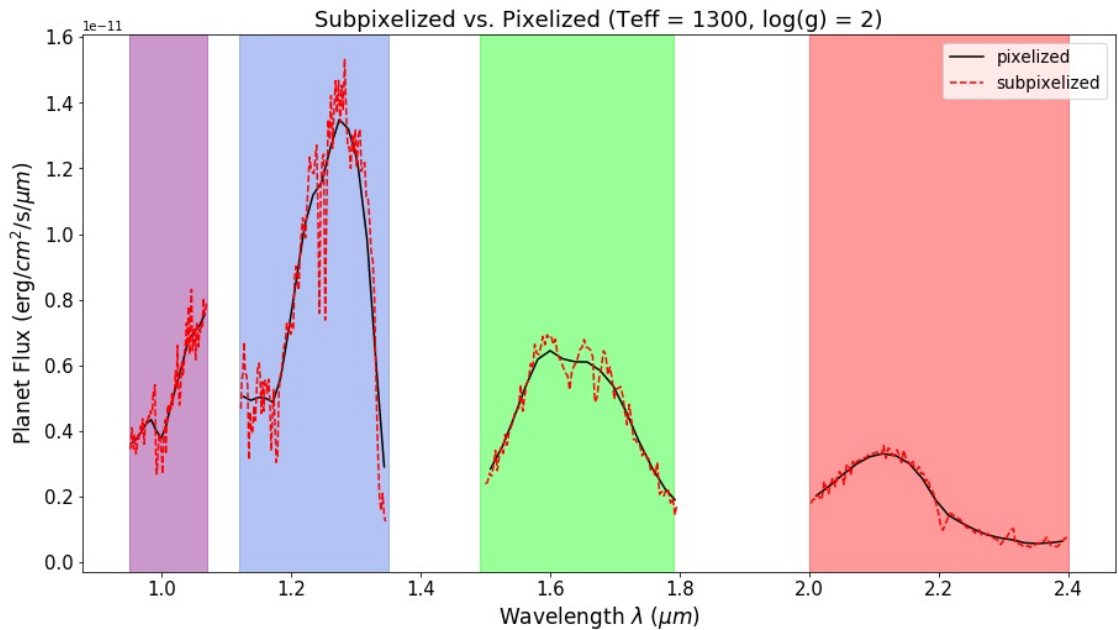

Figure 8. The red dashed curve shows what a model spectrum would look like after the "subpixelization" process where its flux was binned into subpixels. The black solid curve shows what the spectrum would look like after the "subpixelized" spectrum was convolved with the LSF and then re-binned into pixels.

## 3.2 Results

Upon comparing the outcomes of various cases and comparing performances between our base and new simulations, we were able to deduce the optimal instrument parameters and observing scenarios for characterizing exoplanets via direct imaging with GPI. We found that for initial discovery and discriminating between planets and background stars it likely will not be necessary to allocate observing time to all 4 filters, as the H- and J-bands, or even H-band alone, seemed capable of providing sufficient data for constraining temperature. At SNR of 5, the combination of the H- and J-bands demonstrated the best performance amongst cool, mid, and high-temperature planets. Table 4 displays the difference in the minimum and maximum possible temperatures in each confidence interval for a few of the cases we tested.

Table 4. GPI 2.0 Medium Resolution Cases Summary

| Cases | 500 K | | | 1000 K | | | 1300 K | | |
|---|---|---|---|---|---|---|---|---|---|
| | 68% Conf. | 95% Conf. | 99.7% Conf. | 68% Conf. | 95% Conf. | 99.7 Conf. | 68% Conf. | 95% Conf. | 99.7% Conf. |
| H ($t_{exp}$ = 4 hrs) | 5.1 | 21.1 | 35.7 | 7.9 | 30.7 | 49.7 | 31.1 | 113.3 | 192.5 |
| H+Y ($t_{exp}$ = 2 hrs each) | 2.4 | 10.6 | 18.4 | 9.5 | 36.3 | 61.6 | 35.1 | 128.8 | 215 |
| H+J ($t_{exp}$ = 2 hrs each) | 1.2 | 6.9 | 12.2 | 6.3 | 26.1 | 44.3 | 21.5 | 77.7 | 130.4 |
| H+K ($t_{exp}$ = 2 hrs each) | 7.9 | 30.1 | 51.3 | 11.6 | 43.5 | 73.5 | 26.7 | 96.9 | 163.1 |
| Y+J+H+K ($t_{exp}$ = 1 hr each) | 1.8 | 8.7 | 14.8 | 9 | 34.3 | 58 | 24.2 | 87.1 | 146.5 |

Our preliminary results suggest that GPI 2.0 should allow observers to constrain temperature with sufficient accuracy. Although gravity remained largely uncertain across both of our methodologies, the likelihood colormaps we studied demonstrated slightly shorter, narrower ranges which suggests that the new simulation was capable of making estimates with slightly smaller uncertainties.

## 3.3 Low-resolution Broadband Mode Applications

In our original simulation the broadband mode exhibited up to 20% less accuracy than its companion medium resolution mode and the uncertainties in its estimations were 2-3 times larger. Although we observed a significant improvement in accuracy through our new simulation, we concluded that GPI 2.0's low-resolution broadband mode would be better suited for stellar object classification.

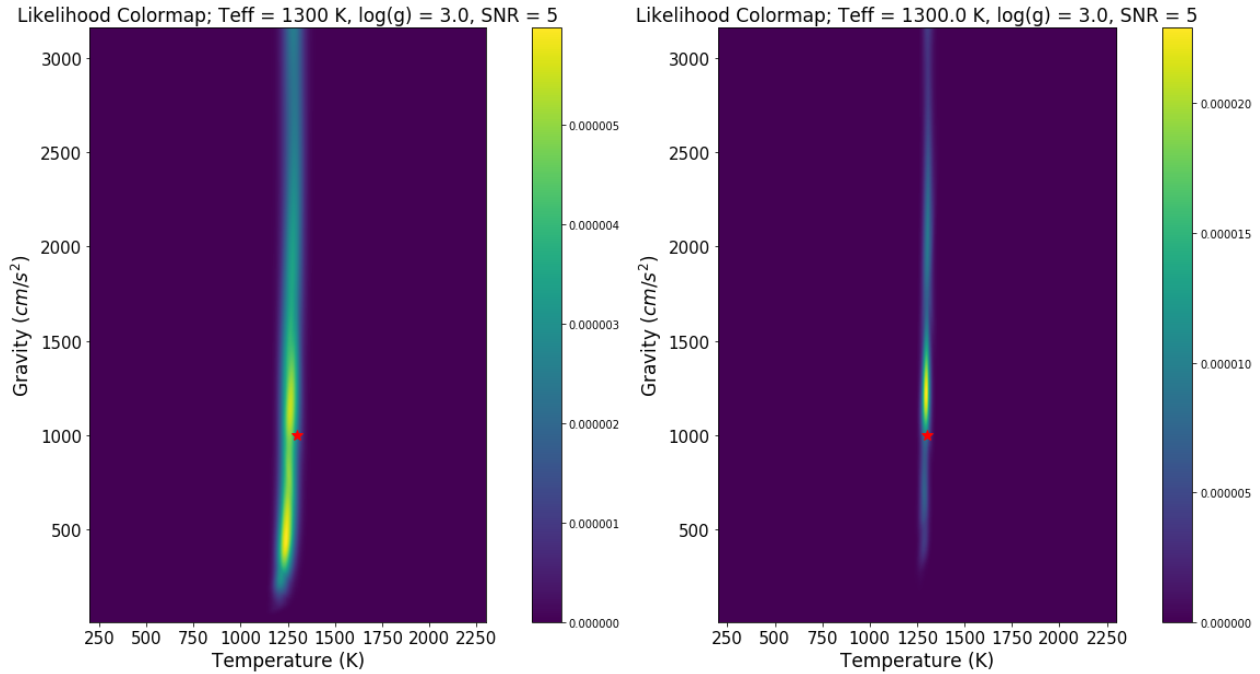

Figure 9. The colormap on the left is from the original simulation. The colormap on the right is from the new simulation and is slightly shorter and narrower than its counterpart.

Background stars are often mistaken for exoplanets (Nielsen et al. 2017[5]), and unfortunately, direct imaging is particularly susceptible to such misidentifications. Ultimately, confirmation requires astrometric orbit, parallax, and/or proper motion measurements, but an initial temperature estimate can help prioritize telescope time for follow-up observations. As such, we sought to determine whether distinguishing background stars from true exoplanet detections would be a realistic application for the new low-resolution broadband mode.

The main question we sought to answer was what the minimum SNR requirement would be for the broadband mode to estimate temperature with enough accuracy to distinguish a real exoplanet from a background star. Figure 10 demonstrates that the fractional error in temperature estimation varies inversely with SNR. Secondly, one can see that with an SNR of approximately 5, all the temperatures in our model grid demonstrated a fractional error of less than 20%. This suggests that future observers using GPI 2.0 will be able to gauge whether an exoplanet candidate is indeed an exoplanet or a background star without having to spend hours of observing time at higher resolutions.

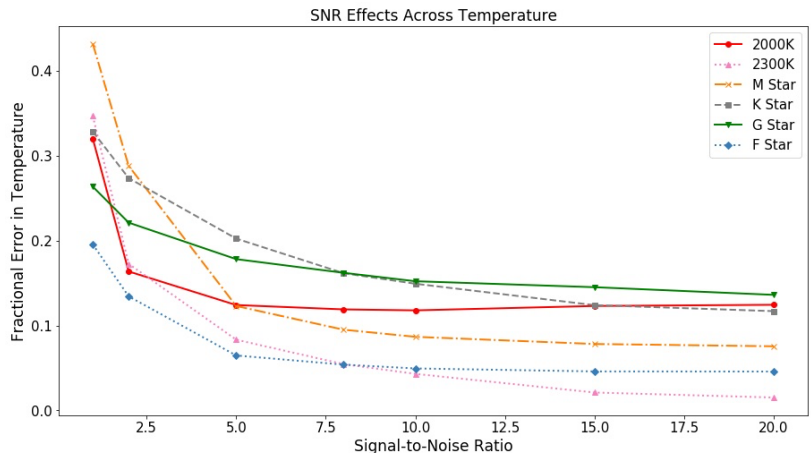

Figure 10. Relation between fractional error in temperature vs. SNR for a handful of different temperature models. There seem to be diminishing marginal returns after SNR ~ 5, suggesting that SNR ~ 5 may be the most ideal SNR requirement.

## 4. CONCLUSION

From the results presented by our simulations, we have deduced that allocating more observing time to a smaller selection of filters may be more efficient than distributing less time between more filters. As shown in Table 4, the best results were obtained from a combination of H- and J-band at an SNR of ~5. Additionally, we observed that the simulated low-resolution broadband mode proved reliable for distinguishing between planet and star spectra. The ability to quickly make the distinction of background stars from planets before additional follow-up observations (e.g., higher spectral resolution or common proper motion) will allow efficient prioritization of telescope time. This will be particularly critical for future space coronagraph observations.

Looking to the future of this project, there are several directions we could take related to improvement and expansion. As asserted by Perrin et al. (2014)[6], "[GPI's] IFS data must be reconstructed into high quality astrometrically and photometrically accurate datacubes in both spectral and polarization modes." One direction for this project could be to model the extraction of spectra with the publicly available data reduction pipeline that meets these needs to see if the pipeline can be improved. For example, fitting models to the raw IFS pixels rather than reconstructed spectra could allow both better modeling of noise and detection of features at higher spectral resolution. Another prospective area of improvement could be developing a more formal statistical Bayesian framework for deciding whether something is a star or a planet – again, particularly critical for space missions. Future analysis will explore the effects of correlated noise, such as that caused by residual speckles (Greco and Brandt 2016[7]), vs the uncorrelated photon and readout noise modeled here. Finally, we are using the same framework to assess the capabilities of Roman-CGI[8] optical spectroscopy to measure gravity and metallicity of known self-luminous planets (Lacy, Burrows 2019[9]). The combination of optical and near-infrared spectroscopy may be particularly powerful for the planets within Roman's reach.